\documentclass[conference]{IEEEtran}
\IEEEoverridecommandlockouts
\usepackage{cite}
\usepackage{amsmath,amssymb,amsfonts}
\usepackage{algorithmic}
\usepackage{graphicx}
\usepackage{textcomp}
\usepackage{xcolor}
\def\BibTeX{{\rm B\kern-.05em{\sc i\kern-.025em b}\kern-.08em
    T\kern-.1667em\lower.7ex\hbox{E}\kern-.125emX}}
\begin{document}

\title{Explainable AI and susceptibility to adversarial attacks: a case study in classification of\\breast ultrasound images}

\author{\IEEEauthorblockN{Hamza Rasaee}
\IEEEauthorblockA{\textit{Electrical and Computer Engineering} \\
\textit{Concordia University}\\
Montreal, Canada \\
h\_rasaee@encs.concordia.ca}

\and
\IEEEauthorblockN{Hassan Rivaz}
\IEEEauthorblockA{\textit{Electrical and Computer Engineering} \\
\textit{Concordia University}\\
Montreal, Canada \\
hrivaz@ece.concordia.ca}
}

\maketitle

\begin{abstract}
Ultrasound is a non-invasive imaging modality that can be conveniently used to  classify suspicious breast nodules and potentially detect the onset of breast cancer. Recently, Convolutional Neural Networks (CNN) techniques have shown promising results in classifying ultrasound images of the breast into benign or malignant. However, CNN inference acts as a black-box model, and as such, its decision-making is not interpretable. Therefore, increasing effort has been dedicated to explaining this process, most notably through GRAD-CAM and other techniques that provide visual explanations into inner workings of CNNs. In addition to interpretation, these methods provide clinically important information, such as identifying the location for biopsy or treatment. In this work, we analyze how adversarial assaults that are practically undetectable may be devised to alter these importance maps dramatically. Furthermore, we will show that this change in the importance maps can come with or without altering the classification result, rendering them even harder to detect. As such, care must be taken when using these importance maps to shed light on the inner workings of deep learning. Finally, we utilize Multi-Task Learning (MTL) and propose a new network based on ResNet-50 to improve the classification accuracies. Our sensitivity and specificity is comparable to the state of the art results.
\end{abstract}

\section{INTRODUCTION}
Breast cancer is a leading cause of death among women worldwide \cite{ref_breath_cancer}. Mammography is usually applied to screen breast cancers as the first imaging modality. However, there are mainly two issues in mammography. First, mammography uses harmful ionizing radiations.  Second, it has a low specificity in distinguishing between dense and cancerous tissues. Therefore ultrasound  has been used as a proper alternative solution for breast cancer screening \cite{ref_1}. Moreover, the ultrasound devices are portable, inexpensive, do not use ionizing radiation and real-time. The advent of recent pocket-size point-of-care ultrasound (POCUS) devices further reduces the costs and improves the portability of this imaging modality. 

Feature-based machine learning methods, such as support vector machine (SVM) and random forest (RF), have been used widely to process and interpret ultrasound images. In the classical machine learning algorithms for ultrasound image processing \cite{ref_z, ref_2, ref_3, ref_4, ref_5}, ultrasound images have been studied as the whole image or divided into different patches to extract key features leading to lesion detection and classification. However, these methods typically require time-consuming preprocessing steps and provide limited sensitivity and specificity. 

The advent of deep learning methods with multiple convolutional and non-linearity layers made it possible for scientists to skip the manual feature extractions and have the deep neural network models to automatically classify the medical images. This exceptional advantage, as well as encouraging results of deep learning methods in the medical images classification task, has attracted a growing interest \cite{ref_6,ref_7,ref_8,ref_9,ref_10,ref_10_1}. In most of these studies\cite{ref_11, ref_12, ref_13}, the main concentration was to distinguish between benign and malignant lesions; although further improvements in deep learning-based methods revealed another point of strength: Multi-Task Learning (MTL). It was shown that training a network with more correlated tasks to predict multiple outputs results in a better performance compared to the same network assigned to only one task \cite{ref_14}. MTL was used to more reliably categorise breast cancer ultrasound images \cite{ref_multi_task_us_pre}.

A major disadvantage of deep learning models is that they are non-transparent and often work as black boxes. This hinders their use in several applications, such as healthcare wherein decisions need to be justified. 
Zhou \textit{et al.} \cite{ref_grad} proposed the concept of utilising Class Activation Mapping (CAM) to describe how a deep learning model predicts the outcomes and discovered that various layers of a CNN behave as object feature extractors. In order to visualize the feature maps, they used the global average pooling (GAP) \cite{ref_cam} layer, and then by combining the feature maps at the layer before the last layer (pre-softmax), they showed a heat map that explains which area of the input image is mapped to the related label. In an follow up work, Selvaraju \textit{et al.} \cite{ref_grad_cam_1} applied an effective generalization of CAM, referred to as GRAD-CAM, that visualizes the input images with high-resolution details to make CNN-based models performance more clear.

The confidence  in a deep learning model is critical, especially in the medical field. Even though explaining a model was made possible by using  GRAD-CAM,  it still  needs to be validated with different data qualities. It is known that ultrasound images can be corrupted by different sources of noise, and their appearance can substantially change by using a different frequency or beamforming approach. These changes can corrupt  the classification results or the GRAD-CAM. Moreover, the predictions of the deep learning networks are susceptible to adversarial attacks \cite{ref_adversarial_1,ref_adversarial_2,ref_adversarial_3}. 
 Ghorbani \textit{et al.} \cite{ref_adversarial_4} applied  adversarial attacks on ImageNet and CIFAR-10 datasets. They revealed that systematic perturbations could cause different interpretations without any modification on the predicted label.

This article presents a MTL model for accurately classifying benign and malignant ultrasound images while simultaneously preparing a mask to localize the lesion. GRAD-CAM is then used to create a feature map to display and describe the model. Finally, the adversarial technique is used to perturb the ultrasound images to evaluate the noise impacts on the specified model and the feature map.
Our contributions can be summarized as below:
\begin{itemize}
    \item Using adversarial perturbations to misclassify the input images.
    \item Using adversarial perturbations to alter the feature maps both with and without changing the classification results.
    \item Multi-class classification of breast ultrasound images with limited training data.
    \item Proposing a new architecture based on ResNet-50 to obtain results similar to the state of the art.
    
\end{itemize} 

\section{METHODS}
Herein, we make two contributions. First, we utilize a CNN for the classification of breast ultrasound images and show the image location responsible for classification. Our classification results are comparable to the state of the art \cite{ref_good_res}. Second, we employ small systematic adversarial perturbations to distort the images such that the classification category does not change, but the location changes. As such, these attacks are hard to detect and can seriously harm subsequent clinical tasks such as biopsy or resection. The ResNet-50 pre-trained on the ImageNet dataset of images has been used to train and test the ultrasound image classification in this study.

The performance and feature map of this model was assessed using GRAD-CAM. We changed the two last layers of the ResNet-50 from fully-connected-1000 to dense-2 with softmax and trained those layers using a public training set consisting of 780 ultrasound images. Finally, the adversarial perturbation is applied to increase the noise of the original image to investigate the predictions with distorted ultrasound images.
\subsection{Dataset}
The first database is retrieved from a public database \cite{ref_breast_1}, and is comprised of breast ultrasound images in PNG format, which was first gathered in 2018. The data were recorded from 600 female patients ranging in age from 25 to 75 years old. There are 780 images in the collection, with an average size of 500x500 pixels. These images are categorized into three groups: 437 benign, 210 malignant, and 133 normal \cite{ref_breast_1}.
The second database contains 250 BMP images of breast cancer, split into 100 benign and 150 malignant images. The images were 72x72 pixels in size, with widths ranging from 57 to 61 pixels and heights ranging from 75 to 199 pixels \cite{ref_breast_2}.
The dataset had two issues for deep learning: first, the image size for feeding the model was varied. Therefore we scaled the photos to 224x224 pixels.

The second issue was the limited amount of training data. As a result, we employed a data generator augmentation approach that included horizontal flip, 5-degree rotation range, height, and a 10\% shift range. Then four images were created for each image. In the end, 70\% of the dataset was used for training, 10\% for validation, and 20\% for testing.  Care was taken to assure there was no data leakage between testing and training.

\subsection{Modified ResNet-50}
\begin{figure}[htbp]
\centerline{\includegraphics[width=90mm,scale=0.5]{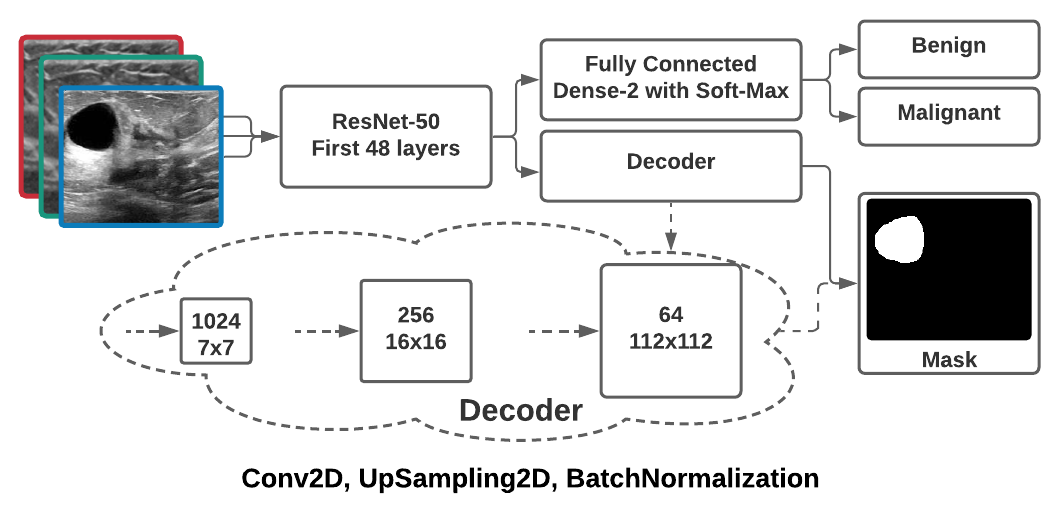}}
\caption{A diagram of our proposed MTL network, showing the top branch for classification and the bottom branch (decoder) for segmentation.}
\label{fig-network}
\end{figure}

In our MTL network for classification and segmentation, we employed the ResNet-50 model as the foundation model. On the one hand, the ResNet-50 model's penultimate layer is a fully-connected-1000 layer that can predict 1000 distinct items. We replaced this layer with a dense-2 (fully-connected-2) and a soft-max layer to forecast our two separate classes. On the other side, we included a decoder after the classifier to produce masks, which aids the classification component in making better predictions (Fig.~\ref{fig-network}). Six sub-boxes were inserted in the encoder box to upsample the ResNet-50 activation layer weights from 7x7x1024 to the mask size (224x224x1). Each sub-box is made up of three smaller sub-boxes: Conv2D with Relu activation, Up Sampling2D(2,2), and Batch Normalization.

\subsection{Gradient-weighted Class Activation Mapping (GRAD-CAM)}
GRAD-CAM was used as a feature map to explain the model.
In general, there are many layers in deep learning to extract features from an image, and as the model becomes more complex,  visual interpretability becomes more important \cite{ref_grad_cam_1}. Only output layer decisions are explained in this paper. Therefore, in this RestNet-50 model, the activation layer includes the majority of visual information linked to the input image before the final layers. The activation layer (layer number 48) is a 7x7 matrix with 2048 channels, so we will end up with a weighted matrix that is the same size as the input image after upsampling 32 times.

\subsection{Adversarial perturbation}
In this section, we use the ``fast gradient sign method'' to inject noise into the input image, which is a common adversarial perturbation approach\cite{ref_adversarial_1}. 
\begin{equation}
\label{f1}
\eta = \epsilon sign(\nabla_xJ(\theta,x,y))
\end{equation}
where $\theta$ is the model's parameter, $x$ is the model's input, $y$ is the predicted label linked to $x$, and $J$ is the trained model's cost function. We can compute the gradient values using backpropagation, and as a consequence, there will be a linear cost function around $\theta$.
We used $\epsilon = 0.1$ with 25 iterations (steps) to ensure that our updated ResNet-50 model was sufficiently disturbed. The input images before and after perturbation are visually identical, but the model incorrectly classifies the data.

\section{Experiments And Results}
In the first step, we primarily aimed to use MTL to create a model that could predict benign and malignant images based on the ResNet-50 model.
The accuracy for single-task learning is 93.61\%, and for MTL is 99.09\%, as indicated in Table.~\ref{tab:acc}. It indicates that it could enhance accuracy by 5.48 percent by utilizing MTL.
 
\begin{table}[htbp]
\caption{Classification results of our MTL-based method.}
\label{tab:acc}
\begin{center}
\begin{tabular}{|c|c|c|}
\hline
\textbf{}&\multicolumn{2}{|c|}{\textbf{Accuracy}} \\
\cline{2-3} 
\textbf{Learning type} & \textbf{\textit{Single-task}}& \textbf{\textit{Multi-task}} \\
\hline
\textbf{Prediction} & \text{\textit{93.61\%}}& \text{\textit{99.09\%}} \\

\hline

\end{tabular}
\label{tab1}
\end{center}
\end{table}

\begin{figure}[htbp]
\includegraphics[width=90mm,scale=0.5]{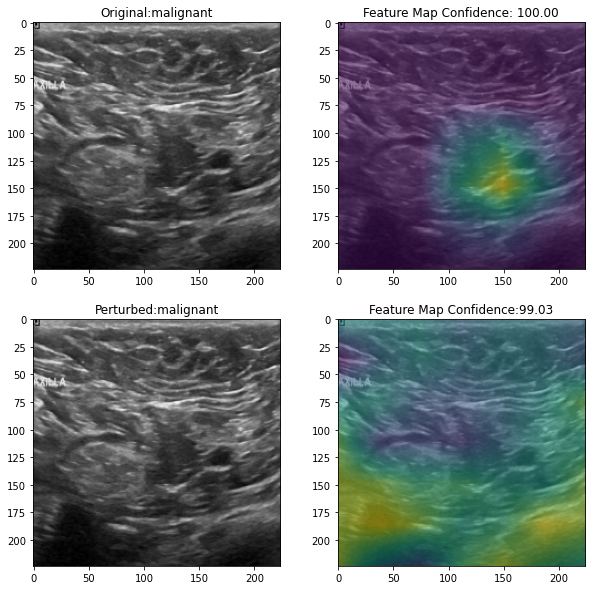}
\caption{The original image on the top left; feature map on the top right; the perturbed image on the bottom left; feature map on the bottom right following adversarial perturbation. The same classification was predicted.}
\label{fig-same}
\end{figure}

\begin{figure}[htbp]
\centerline{\includegraphics[width=90mm,scale=0.5]{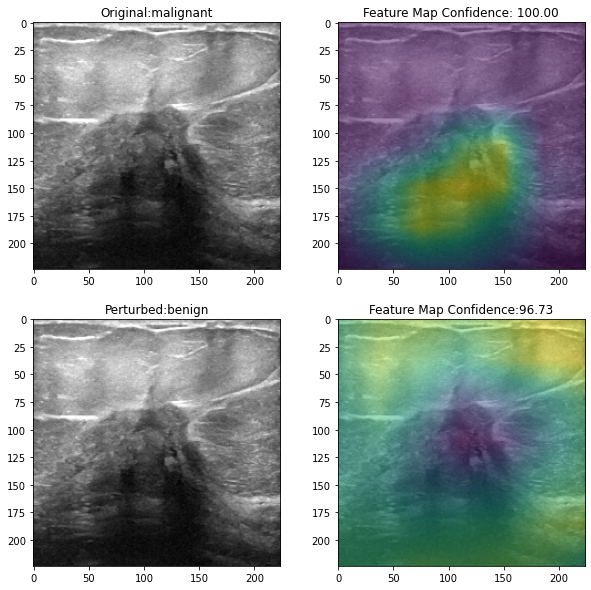}}
\caption{The original image on the top left; feature map on the top right; the perturbed image on the bottom left; feature map on the bottom right following adversarial perturbation. A different classification was predicted.}
\label{fig-diff}
\end{figure}

\begin{figure}[htbp]
\centerline{\includegraphics[width=90mm,scale=0.5]{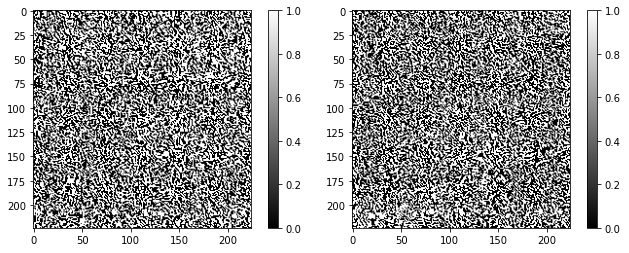}}
\caption{The left and right images show the differences between images before and after adversarial attacks of Figs.~\ref{fig-same} and \ref{fig-diff}, respectively. Note that the range in this image is between 0 to 1. The range in B-mode intensities in Figs.~\ref{fig-same} and \ref{fig-diff} is 0 to 255. }
\label{difference-image}
\end{figure}

After the disturbance, there are primarily two sorts of impact on the images. The adversarial perturbation changes the feature map, as seen in Fig.~\ref{fig-same}. As a result, the feature map weights move from the center to the bottom-left of the image. Yet, the predictions on the original and perturbed images are the same with extremely high confidence. The adversarial perturbation on the image deceives the network in the second scenario. Fig.~\ref{fig-diff} indicates that the model categorized the image as malignant with 100\% confidence before perturbation, while the model wrongly predicts benign with high confidence after the adversarial perturbation. The difference between the original and perturbed images in Figs.~\ref{fig-same} and ~\ref{fig-diff} is shown in Fig.~\ref{difference-image}. The maximum value of the difference is 1, whereas the maximum value of B-mode images is 255. In other words, the changes in the B-mode image caused by the adversarial attack is very small.

\section{CONCLUSIONS}
Even though the designed MTL based on the ResNet-50 model can identify the input images as benign or malignant with a high accuracy (99.09\%), the images before and after our modest adversarial perturbations seem virtually the same. While adversarial assaults on CNNs are well-known, our findings demonstrate that interpretations of breast ultrasound images are also subject to similar attacks.

\section*{Acknowledgment}
We acknowledge the support of the Natural Sciences and Engineering Research Council of Canada (NSERC).

\end{document}